\documentclass[prl,twocolumn,aps]{revtex4-1}
\usepackage{bm,graphicx,color,enumerate}
\usepackage{amsmath,amssymb,amsfonts}

\begin{document}

\title{On the velocity-strengthening behavior of dry friction}

\author{Yohai Bar-Sinai$^{1}$, Robert Spatschek$^{2}$, Efim A. Brener$^{1,3}$ and Eran Bouchbinder$^{1}$}
\affiliation{
$^1$ Chemical Physics Department, Weizmann Institute of Science, Rehovot 76100, Israel\\
$^2$ Max-Planck-Institut f\"ur Eisenforschung GmbH, D-40237 D\"usseldorf, Germany\\
$^3$Peter Gr\"unberg Institut, Forschungszentrum J\"ulich, D-52425 J\"ulich, Germany
}

\begin{abstract}
The onset of frictional instabilities, e.g. earthquakes nucleation, is intimately related to velocity-weakening friction, in which the frictional resistance of interfaces decreases with increasing slip velocity. While this frictional response has been studied extensively, less attention has been given to steady-state velocity-strengthening friction, in spite of its potential importance for various aspects of frictional phenomena such as the propagation speed of interfacial rupture fronts and the amount of stored energy released by them. In this note we suggest that a crossover from steady-state velocity-weakening friction at small slip velocities to steady-state velocity-strengthening friction at higher velocities might be a generic feature of dry friction. We further argue that while thermally activated rheology naturally gives rise to logarithmic steady-state velocity-strengthening friction, a crossover to stronger-than-logarithmic strengthening might take place at higher slip velocities, possibly accompanied by a change in the dominant dissipation mechanism. We sketch a few physical mechanisms that may account for the crossover to stronger-than-logarithmic steady-state velocity-strengthening and compile a rather extensive set of experimental data available in the literature, lending support to these ideas.
\end{abstract}

\maketitle
\section{Introduction}

Understanding the constitutive behavior of dry frictional interfaces has far-reaching implications for a broad range of phenomena and scientific disciplines \citep{Armstrong-Helouvry1994,Scholz1998, Marone1998, Olsson1998, Persson2000, Scholz2002, Urbakh2004, Baumberger2006, Ben-Zion2008, Kawamura2012, Vanossi2013, Ohnaka2013}. It is well-established that the onset of frictional instabilities, which might lead to interfacial failure (e.g. earthquakes), is intimately related to weakening effects, i.e. the reduction of frictional resistance with increasing slip displacement or slip velocity \citep{Rice1983}. In particular, when the slip velocity $v$ is regarded as a basic frictional control variable, the variation of the steady-state frictional resistance with $v$ is of great importance. Naturally, steady-state velocity-weakening (denoted hereafter as SVW) friction has been studied extensively. On the other hand, less attention has been given to steady-state velocity-strengthening (denoted hereafter as SVS) friction, in which the steady-state frictional resistance increases with increasing $v$. The existence of SVS might affect, for example, the propagation speed of rupture fronts, their propagation distance and possibly the amount of stored energy released by them. Our goal in this note is to discuss SVS friction, its functional form and possible physical origins, and to point out direct evidence for its existence based on experimental data available in the literature.

The interface between two macroscopic bodies in dry frictional contact is typically composed of an ensemble of contact asperities whose total area $A_r$ is orders of magnitude smaller than the nominal contact area $A_n$. The real contact area typically depends on the time elapsed since a contact was formed, i.e. on the contact's ``age'' (or ``maturity'') typically quantified by a state variable of time dimension $\phi$ \citep{Dieterich1978, Ruina1983, Dieterich1994, Rice2001, Baumberger2006}, an idea that dates back at least to \citet{Rabinowicz1951}. The frictional stress (resistance) $\tau$ is proportional to $A_r(\phi)$ \citep{Bowden1964}. The proportionality factor depends on the slip velocity $v$ and possibly on a set of internal state variables which we schematically denote by $\theta$, and can be interpreted as the shear strength $\sigma_s(\theta, v)$ (related to the plastic flow of contact asperities) \citep{Bowden1964, Baumberger2006}. This contribution to the frictional resistance is rheological in nature. Putting the two together, one can write the frictional stress (resistance) as \citep{Bowden1964, Baumberger2006}
\begin{equation}
\label{nonsteady}
\tau(\phi,\theta, v) = \frac{A_r(\phi)\,\sigma_s(\theta, v)}{A_n} \ .
\end{equation}
During steady-state sliding at a velocity $v$, the internal state variables attain unique values $\phi(v)$ and $\theta(v)$. Therefore, under steady-state conditions the frictional stress $\tau^{ss}(v)$ takes the form
\begin{equation}
\label{steady}
\tau^{ss}(v) = \frac{A_r[\phi(v)]\,\sigma_s[\theta(v),v]}{A_n} \ .
\end{equation}

In this note we focus on the variation of $\tau^{ss}(v)$ with $v$, and in particular on the sign of $\partial_v \tau^{ss}$, its dependence on $v$ and its functional form. $\partial_v \tau^{ss}\!<\!0$, i.e. SVW friction, is known to facilitate unstable accelerating slip and frictional instabilities \citep{Ruina1983, Scholz1998}. On the other hand, $\partial_v \tau^{ss}\!>\!0$, i.e. SVS friction, might promote stable slip, limit the propagation speed of interfacial rupture fronts and affect the magnitude of slip events \citep{Weeks1993, Kato2003, Shibazaki2003, Bouchbinder2011, BarSinai2012, Hawthorne2013, Bar-Sinai2013pre}, limit the seismogenic zone \citep{Marone1988} and affect earthquake afterslip and stress drops \citep{Marone1991}. In what follows we suggest that SVS friction, $\partial_v \tau^{ss}\!>\!0$, generically emerges in dry friction over some range of slip velocities, discuss its possible physical origins and the available experimental evidence for its existence.

\section{Real Contact Area Aging and its Saturation During Sliding}

Equation (\ref{steady}) suggests that the steady-state frictional stress $\tau^{ss}(v)$ is a product of a steady-state real contact of area contribution $A_r^{ss}(v)\!\equiv\!A_r[\phi(v)]$ and a rheological contribution $\sigma^{ss}_s(v)\!\equiv\!\sigma_s[\theta(v),v]$. As the latter is assumed to be an increasing function of $v$, $\partial_v \sigma^{ss}_s\!>\!0$ (to be discussed later), we focus first on $A_r^{ss}(v)$.

The physical argument we present has already appeared in the work of Baumberger, Caroli and coworkers \citep{Baumberger1999, Bureau2002, Baumberger2006}, in a slightly different form in \citet{Putelat2011} and in \citet{BarSinai2012}. It is repeated here briefly for completeness. The starting point is the non-steady behavior of $A_r(\phi)$ in the absence of slip, $v\!=\!0$. In this case, $\phi$ grows linearly with the time $t$ elapsed since the frictional interface was formed (when the bodies under consideration brought into frictional contact) or since previous slip halted, $\phi\!=\!t$.

It is well-established that under these conditions the real contact area undergoes logarithmic aging, i.e. $A_r(t)$ increases logarithmically with $t$, a behavior observed in many materials \citep{Dieterich1972, Ruina1983, Dieterich1994, Marone1998, Berthoud1999, Nakatani2001, Nakatani2006, Baumberger2006, Ben-David2010b}. Specifically, the time evolution of $A_r(t)$ (normalized here by $A_n$) takes the following form
\begin{equation}
\label{wrong aging}
\frac{A_r(t)}{A_n}\!=\! \frac{\sigma}{\sigma_{\hbox{\tiny H}}}\left[1 + b\log(t/\phi^*)\right] \ ,
\end{equation}
where $b$ and $\phi^*$ are positive constants \citep{Ruina1983}, $\sigma$ is the normal stress and $\sigma_{\hbox{\tiny H}}$ is the material's hardness. This expression, however, cannot be valid for arbitrarily short times as it becomes singular for $t\!\to\!0$, which is of course unphysical. This simply means that $A_r(t)$ actually takes the form
\begin{equation}
\label{right aging}
\frac{A_r(t)}{A_n}\!=\!\frac{\sigma}{\sigma_{\hbox{\tiny H}}}\left[1 + b\log(1 + t/\phi^*)\right] \ ,
\end{equation}
where $\phi^*$ can be interpreted as a typical cutoff time scale for the onset of logarithmic aging. Equation (\ref{wrong aging}) provides a good approximation for Eq. (\ref{right aging}) when $t\!\gg\!\phi^*$, but completely fails in the opposite limit, $t\!\ll\!\phi^*$. The last equation, and specifically the short-time cutoff, has been proposed by \citet{Dieterich1978}, has been verified experimentally \citep{Marone1998Nature, Nakatani2006, Ben-David2010b} and has been studied theoretically \citep{Estrin1996, Nakatani2006, Putelat2011}.

While this might appear as a somewhat academic discussion of a short time regularization of the logarithmic aging formula, and indeed it is almost always overlooked, this is not the case. To see the relevance of this short time regularization for our purposes here, we should consider steady sliding at a velocity $v$. Since $\phi$ quantifies the age of the real contact it must be a decreasing function of $v$ (the ``lifetime'' of a contact asperity is shorter the higher the slip velocity, i.e. ``rejuvenation''). It is well established that under steady-state conditions \citep{Dieterich1978, Baumberger1999, Baumberger2006}
\begin{equation}
\label{phiss}
\phi = D/v \ ,
\end{equation}
where $D$ is a typical slip distance (usually related to the contact asperities size) \citep{Marone1998, Baumberger2006}. Therefore, under steady-state sliding conditions the real contact area takes the form
\begin{equation}
\label{Ass}
\frac{A_r^{ss}(v)}{A_n}\!=\!\frac{\sigma}{\sigma_{\hbox{\tiny H}}}\left[1 + b\log\left(1 + \frac{D}{v\phi^*}\right)\right] \ .
\end{equation}
This implies that $A_r^{ss}(v)$ decreases logarithmically with increasing $v$ for $v\!\lesssim\! D/\phi^*$ \citep{Teufel1978} and that it approaches a constant (saturates) for $v\!\gg\!D/\phi^*$. Therefore, if indeed the rheological contribution to the steady-state frictional resistance increases with $v$, $\partial_v \sigma_s^{ss}\!>\!0$, we conclude that {\em irrespective} of the precise form of $\sigma_s^{ss}(v)$ we expect $\partial_v \tau^{ss}\!>\!0$ for $v\!\gg\!D/\phi^*$. This is an important observation.

\section{Logarithmic Steady-State Velocity- \newline Strengthening Friction}

The last section concluded with the observation that the steady-state frictional stress (resistance) $\tau^{ss}(v)$ is expected to become velocity-strengthening above a certain slip velocity $\sim\!D/\phi^*$ due to the saturation of the real contact area, assuming $\partial_v \sigma_s^{ss}\!>\!0$. Our goal in this section, and the subsequent one, is to discuss the latter.

The standard approach to the velocity dependence of the rheological part of the frictional stress (``shear strength'') is to attribute it to thermal activation \citep{Baumberger1999, Rice2001, Baumberger2006, Putelat2011}. We briefly repeat the argument here as it sets the stage for what will follow. The starting point is to treat the real contact of area $A_r(\phi)$ as fixed, to neglect any rheological internal variables $\theta$ and to assume that $v$ is a result of a stress-biased thermally activated process such that
\begin{equation}
\label{v} v = v_0\left(\exp\left[-\frac{\Delta(\tau)}{k_B T}\right]-\exp\left[-\frac{\Delta(-\tau)}{k_B T}\right]
\right) \ .
\end{equation}
Here $v_0$ is a reference velocity scale related to a basic attempt rate and an intrinsic length scale, $\Delta(\tau)$ is a stress-biased activation barrier, $k_B$ is Boltzman's constant and $T$ is the temperature. The second exponential appears in order to account for backwards transitions (implying a proper $\tau\!\to\!-\tau$ symmetry and consistency with the second law of thermodynamics).

The stress-biased activation barrier is assumed to take the form
\begin{equation}
\label{linear_bias} \Delta(\tau)=E_0-\Omega\,\tau^{loc}(\tau) \ ,
\end{equation}
where $E_0$ is the bare energy barrier, $\Omega$ is the activation volume (typically much larger than atomic volumes, i.e. corresponding to a collective multi-atom process \citep{Rice2001, Baumberger2006}) and $\tau^{loc}(\tau)\!=\!A_n\tau/A_r$
is the local stress at the asperity level. $E_0$ is the energy barrier in equilibrium, where forward and backward thermally activated transitions are equally likely and $v\!=\!0$. The application of a stress $\tau^{loc}$ favors transitions in its direction over transitions in the opposite direction, giving rise to $v\!\ne\!0$.
Note that the local asperity stress $\tau^{loc}$ is significantly enhanced compared to the macroscopic stress $\tau$ by a large factor $A_n/A_r\!\gg\!1$.
Therefore, we can rewrite Eq. (\ref{v}) as
\begin{equation}
 v=2\,v_0\exp\left[-\frac{E_0}{k_B T}\right]\sinh\left(\frac{A_n \Omega\,\tau}{A_r k_B T}\right) \ , \label{vpl}
\end{equation}
which can be inverted in favor of the stress to read (putting back the $\phi$ dependence of $A_r$)
\begin{equation}
\tau(\phi,v) = \frac{k_B T A_r(\phi)}{\Omega\,A_n}\sinh^{-1}\left(\frac{v}{2\,v_0} \exp\left[\frac{E_0}{k_B T}\right]\right)  \ .
\label{tau}
\end{equation}

Finally, since $E_0$ is typically much larger than $k_B T$, we can treat the argument of the inverse $\sinh$-function as large for all $v$'s of interest and approximate $\sinh^{-1}(x)\!\simeq\!\log(2x)$, yielding
\begin{equation}
\tau(\phi,v) = \frac{A_r(\phi)}{A_n}\left[\frac{E_0}{\Omega} + \frac{k_B T}{\Omega}\log\left(\frac{v}{v_0}\right)\right] \ .
\label{tau1}
\end{equation}

The frictional stress in Eq. (\ref{tau1}) takes the form assumed in Eq. (\ref{nonsteady}) and $\sigma_s(v)$ can be readily identified. Equation (\ref{tau1}) predicts that the instantaneous response (i.e. faster than the typical evolution time of $\phi$) of the frictional stress to slip velocity ``jumps'' would be logarithmic in the ratio between the final and the initial $v$'s. This logarithmic ``direct effect'' \citep{Marone1998, Baumberger2006} has been observed for many materials in a wide range of slip velocities $v$, but typically not larger than a few hundreds of $\mu$m/sec.

What are the implications of Eq. (\ref{tau1}) for the steady-state frictional stress $\tau^{ss}(v)$? To answer this question one should substitute $A_r^{ss}(v)$ of Eq. (\ref{Ass}) for $A_r(\phi)$ in Eq. (\ref{tau1}) to obtain
\begin{eqnarray}
\frac{\tau^{ss}(v)}{\sigma} &\equiv& \mu^{ss}(v) = f_0+\alpha \log\left(\frac{v}{v_0}\right)+\beta\log\left(1 + \frac{D}{v \phi^*}\right) \nonumber\\
&+&\frac{\alpha\beta}{f_0}\log\left(\frac{v}{v_0}\right)\log\left(1 + \frac{D}{v \phi^*}\right)\ ,\label{eq:RSF}
\end{eqnarray}
where
\begin{equation}
\alpha \equiv \frac{k_B T}{\sigma_{\hbox{\tiny H}} \Omega}, \quad \beta \equiv \frac{E_0 b}{\sigma_{\hbox{\tiny H}}\Omega}, \quad f_0\equiv \frac{E_0 }{\sigma_{\hbox{\tiny H}}\Omega} = \frac{\beta}{b}\ .\label{eq:rsf^defs}
\end{equation}
As typically the energy scale $\sigma_{\hbox{\tiny H}} \Omega$ is much larger than $k_B T$, we expect $\alpha\!\ll\!1$, which is confirmed by numerous experiments \citep[for example]{Kilgore1993, Heslot1994, Nakatani2001}. The aging coefficient $b$ is typically much smaller than unity (for many materials it is of the order of $10^{-2}$ \citep{Baumberger2006}) and we expect $E_0\!\lesssim\!\sigma_{\hbox{\tiny H}} \Omega$. This implies that $f_0$, which sets the overall magnitude of the friction coefficient, is roughly of order unity (as is widely observed), and that $\beta\!\ll\!1$. These estimates suggest that the last term in Eq. (\ref{eq:RSF}), which is proportional to $\alpha\beta/f_0$, is small compared to the other $v$-dependent terms in this equation and hence will be neglected hereafter (unless otherwise stated).

Consider then relatively small slip velocities that satisfy $v\!\ll\!D/\phi^*$. In this case the $v$-dependence of $\mu^{ss}(v)$ is logarithmic and we have
\begin{equation}
\frac{\partial\mu^{ss}(v)}{\partial\log{v}} = \alpha - \beta = \frac{k_B T}{\sigma_{\hbox{\tiny H}} \Omega}\left(1 - \frac{E_0 b}{k_B T} \right) \ .\label{logarithmicder}
\end{equation}
The sign of the last expression, which is controlled by the relative magnitudes of $b$ and $k_B T/E_0$ (or alternatively of $\alpha$ and $\beta$), determines whether friction is velocity-weakening or velocity-strengthening in this range of slip velocities. In particular, for $b\!>\!k_B T/E_0$ friction is velocity-weakening and for $b\!<\!k_B T/E_0$ it is velocity-strengthening. In the former case, steady-state friction is velocity-weakening for $v\!\ll\!D/\phi^*$ and then it crosses over to velocity-strengthening behavior for $v\!\gg\!D/\phi^*$, when $A_r(\phi)$ approaches a constant. In the latter case, friction is logarithmically velocity-strengthening for slip velocities both below and above $D/\phi^*$, but with different pre-factors in each regime. Indeed, in some systems such as clay-rich fault gauge layers, velocity-strengthening friction is the rule rather than the exception \citep{Noda2009, Ikari2009, Ikari2013}. The most important implication then of Eqs. (\ref{Ass}) and (\ref{tau1}), for our purposes here, is that they predict that steady-state friction is logarithmically velocity-strengthening for $v\!\gg\!D/\phi^*$. This is often overlooked in parts of the literature, but see \citet{Bureau2002, Baumberger2006}.

In Fig. \ref{exp_evidence}a-b we present examples from the available literature in which logarithmic SVW friction crosses over to logarithmic SVS behavior at some slip velocity $v_m$ (where the curve reaches a minimum). The data in Fig. \ref{exp_evidence}c also exhibit logarithmic SVS.

\section{Stronger-than-Logarithmic Steady-State Friction}
\label{sec:rehology}
As discussed above, a simple thermal activation model predicts the existence of logarithmic SVS friction above a slip velocity $\sim\!D/\phi^*$. This model assumes a single activation barrier and a single attempt rate; many of the materials of interest, however, are disordered and hence a distribution of activation barriers and time scales might be relevant. Moreover, the linear dependence of the activation barrier on $\tau$ in Eq. (\ref{linear_bias}) might not be always valid. In spite of these simplifications and possible limitations, we adopt this framework and ask
whether the logarithmic velocity-strengthening behavior might break down at some point.

Obviously, the simple thermal activation picture breaks down when the stress-biased barrier in Eq. (\ref{linear_bias}) becomes comparable to $k_B T$. Alternatively, the breakdown occurs when the slip velocity $v$ is not much smaller than $v_0$ in Eq. (\ref{v}). There is, however, no easy way to independently estimate both $v_0$ and $E_0$, which are coupled in Eq. (\ref{tau1}). Moreover, as $E_0$ appears in the exponential, small variations in it can be compensated by a huge variation in $v_0$, which suggests a large uncertainty in the latter. Nevertheless, some arguments for independently estimating $v_0$ have appeared in the literature. For example, \citet{Rice2001} estimated $v_0$ for rocks (quartzite and granite) to be in the mm/sec range, which sets an upper bound for the validity of the thermal activation process and hence for logarithmic velocity-strengthening (assuming, for the moment, that $v_0\!>\!D/\phi^*$).

What happens for larger slip velocities, $v\!\gtrsim\!v_0$, when thermal activation breaks down? While we suspect that the answer might be material-specific, we believe that rather generically steady-state friction remains velocity-strengthening in this regime (at least until thermal weakening possibly intervenes, see Section \ref{summary}), with a functional dependence which is typically {\em stronger than logarithmic}. Our basic argument is that the breakdown of the thermal activation process should also signal a change in the dominant energy dissipation mechanism associated with frictional dynamics. Since logarithmic velocity-strengthening is usually intimately linked to thermal activation, we see no reason for other dissipative processes to give rise to such a weak (i.e. logarithmic) dependence on $v$ and hence expect the dependence to be stronger than logarithmic.

\begin{figure*}
\centering
 \includegraphics[width=\textwidth]{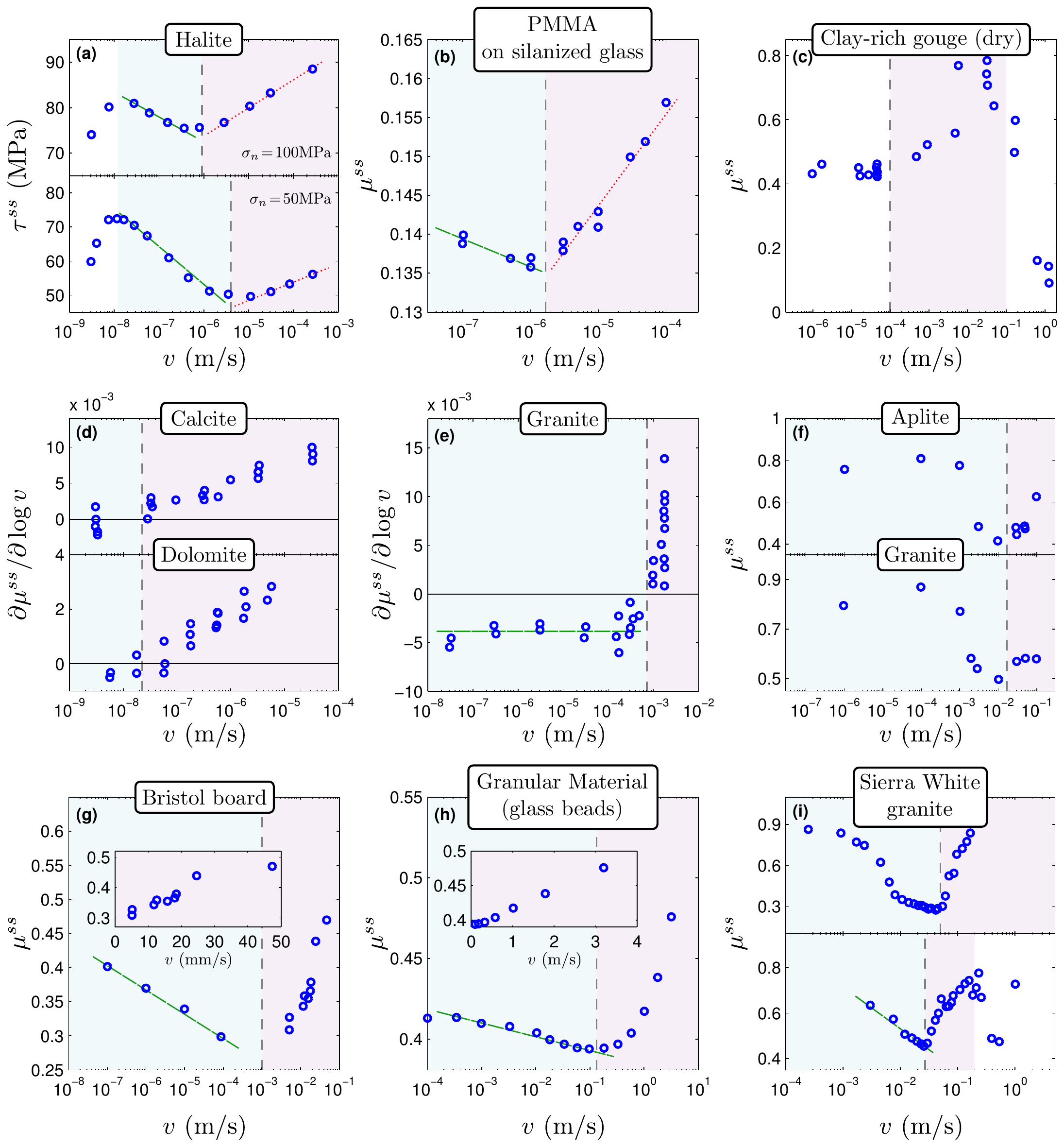}
 \caption{\textbf{Experimental observations of steady-state velocity-strengthening friction in various materials.}  In each panel the region corresponding to velocity-weakening friction (light blue background) is separated by a vertical dashed line from the subsequent region corresponding to velocity-strengthening friction (light purple background). White background corresponds to either velocity-independent friction or a region that is not discussed in this paper. In all of the panels the horizontal axis is $\log{v}$ and the vertical panel is either $\tau^{ss}$ or $\mu^{ss}$, with the exception of panels (d-e), where the vertical axis corresponds to the logarithmic derivative of the friction coefficient, $\partial\mu^{ss}/\partial\log{v}$. Green dashed lines mark logarithmic SVW, while red dotted lines mark logarithmic SVS. The figure is described in detail in Section \ref{sec:evidence}.}
 \label{exp_evidence}
\end{figure*}

While we do not aim here at developing a detailed model of the crossover to stronger-than-logarithmic steady-state friction, we would like to sketch a few possible physical scenarios that might give rise to such a behavior. We first consider the possibility that logarithmic SVS friction crosses over to a linear behavior in which $\tau^{ss}\!\propto\! v$ (see, for example, Fig. \ref{exp_evidence}g and Fig. 15 in \citet{Baumberger1999}). This viscous-friction behavior might emerge as a standard viscous process obtained through linearization of a different thermally activated process, characterized by an activation volume significantly smaller than $\Omega$. That is, if at high slip velocities (hence higher stresses) the physics of frictional dissipation changes such that the activation volume $\Omega$ decreases from a multi-atom/super-molecular value to an atomic/molecular volume, the thermal activation formula of Eq. (\ref{tau}) remains valid, but now $\Omega \tau^{loc}\!\ll\!k_B T$ (recall that $\tau^{loc}\!=\!A_n\tau/A_r$) and linearization leading to $\tau\!\propto\!v$ is sensible (\textit{Caroli}, private communication, 2013).

Another mechanism that may lead a crossover to a linear viscous behavior is well-known in the context of dislocation mechanics \citep{Hirth1967}. In this case, at relatively small mean dislocation velocities and applied stresses, dislocation motion is thermally activated with the barriers determined by local obstacles of various types and the Peierls lattice potential. At higher velocities and stresses, interactions with phonons and electrons control dislocation motion, leading to a linear drag-like relation between the stress and the velocity \citep{Kumar1969,Zerilli1992}.

A crossover from a thermally activated regime at low slip velocities to a non-thermally activated regime at higher slip velocities has been briefly discussed in \citet{Baumberger2006}. The idea there was that plastic rearrangements at contact asperities give rise to mesoscopic stress fields that perturb nearby regions. The accumulated effect of these random spatiotemporal perturbations, originating from various plastic rearrangements taking place at different locations and times, can be regarded as a dynamical/mechanical noise, which acts in parallel to the ordinary thermal noise. This dynamical/mechanical noise becomes more intense as $v$ increases and eventually takes over the thermal noise. While the precise functional form of the velocity-strengthening frictional response associated with the dynamical/mechanical noise-controlled regime has not been discussed, it was implied that it is stronger than the logarithmic dependence associated with the thermal noise-controlled regime (cf. Fig. 17 in \citet{Baumberger2006}).

Another physical scenario for the velocity-strengthening frictional response in the non-thermally activated regime might be based on applying Bagnold's scaling arguments, originally developed in the context of dense granular flows \citep{Bagnold1956}, to atomic/molecular systems \citep{Langer2007}. The idea is that when thermal activation is irrelevant, the system has no characteristic energy scale and flow rates are controlled by collisions between hard-core-like objects, where the detailed molecular interactions are not playing a central role. In this case, the frictional stress $\tau$ is proportional to the product of the momentum transfer per collision and the rate of collisions, both linear in $v$, leading to $\tau\!\propto\!v^2$. While the application of Bagnold's scaling arguments to atomic/molecular systems might be questionable, our goal here is just to highlight another known mechanism for a velocity-strengthening type of response.

In strictly athermal frictional interfaces, e.g. frictional interfaces composed of granular materials such as fault gouge, where the elementary units are macroscopic and no thermal motion takes place, we do not expect logarithmic velocity-strengthening to emerge. In this case, one might expect logarithmic velocity-weakening friction at small slip velocities, due to logarithmic aging of the contacts between grains, to cross over to a stronger-than-logarithmic velocity-strengthening friction associated with nonlinear plastic rheology. This is precisely what has been observed and discussed very recently in \citet{Kuwano2013} (see also Fig. \ref{exp_evidence}h).

Finally, we note that there might exist additional strengthening mechanisms, that go beyond the somewhat idealized multi-contact interfaces picture discussed above, associated with wear and gouge accumulation. For example, the nonmonotonic steady-state friction of Sierra White granite shown in Fig. \ref{exp_evidence}i, including both velocity-weakening and velocity-strengthening behaviors, has been suggested to be directly controlled by the rate of wear formation \citep{Reches2010}. The role of comminution as a dissipative mechanism that leads to frictional strengthening has been extensively discussed and demonstrated in \citet{Spray2005, Spray2010}.

This qualitative discussion of possible physical mechanisms that might give rise to stronger-than-logarithmic velocity-strengthening friction when thermal activation breaks down is only meant to show that such mechanisms are conceivable. The generic picture that emerges is that in thermal systems, when $v_0$ is sufficiently larger than $D/\phi^*$ and $b\!>\!k_B T/E_0$, we expect logarithmic SVW friction to cross over to logarithmic SVS friction at slip velocities $v\!\gtrsim\!D/\phi^*$, which in turn crosses over to stronger-than-logarithmic velocity-strengthening friction at slip velocities $v\!\gtrsim\!v_0$. When $v_0\!<\!D/\phi^*$, we expect logarithmic SVS friction to cross over to stronger-than-logarithmic velocity-strengthening friction, not due to the saturation of the real contact area, but rather because stronger-than-logarithmic strengthening takes over logarithmic weakening. This is also the case for strictly athermal frictional interfaces.

To conclude this section, we note that while the discussion above -- starting with Eq. (\ref{nonsteady}) -- has focussed primarily on frictional interfaces whose contact asperities deform plastically, a similar picture has been discussed by Byerlee in the context of frictional interfaces composed of geological materials governed by brittle fracture of asperities \citep{Byerlee1967a}. Moreover, we would like to draw the readers' attention to the work of Estrin and Br\'echet \citep{Estrin1996}, who seem to discuss somewhat related ideas. Finally, it is important to mention that in the context of lubrication, i.e. frictional interfaces that contain fluids, the generic steady-state dry friction curve $\tau^{ss}(v)$ discussed here is the standard known as the ``Stribeck curve'' \citep{stribeck1903,Olsson1998}. In this curve, solid contact dominates at small slip velocities and hydrodynamic viscous friction dominates at high slip velocities, with a mixed regime in between, where friction goes through a minimum.

\section{Experimental Evidence}
\begin{figure*}
 \includegraphics[width=\textwidth]{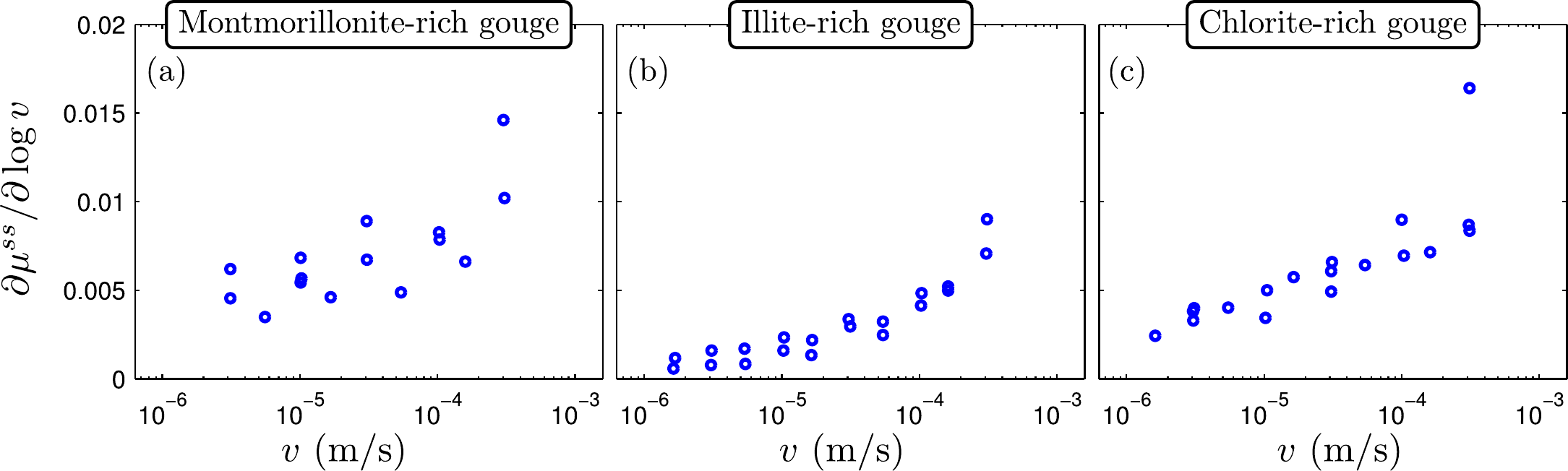}
 \caption{Logarithmic derivative of the steady-state friction coefficient for various saturated clay-rich fault gouges. Data extracted from Fig. 9 of \citep{Ikari2009}, where $\alpha\!-\!\beta$ (which is identical to  $\partial\mu^{ss}/\partial\log{v}$) were reported. It is seen that these systems are velocity-strengthening throughout the explored velocity range. }
 \label{clay}
\end{figure*}
\label{sec:evidence}
To test the physical picture described above, we have searched the available literature, looking for steady-state friction experiments that go up to sufficiently high slip velocities. While these experiments are not easy to perform, and sometimes require to employ different experimental techniques in different ranges of slip velocities, we have been able to trace quite a few examples that lend support to the proposed picture. All of the panels of Fig. \ref{exp_evidence}, which span a rather wide range of materials, clearly exhibit SVS. Note that we classify a SVS behavior as ``stronger-than-logarithmic'' whenever it cannot be reasonably described by a $\log{v}$ variation, with a sensibly small pre-factor. Here we provide additional information about the data presented in the figure:
\begin{enumerate}[(a)]
 \item $\tau^{ss}(v)$ for a pre-cut fault in halite in a triaxial apparatus, data extracted from Fig. 3 of \citet{Shimamoto1986}. The lower panel corresponds to a normal stress of $\sigma\!=\!50$ MPa and the upper one to $\sigma\!=\!100$ MPa. A crossover from logarithmic SVW (marked by a negative slope green dashed line) to logarithmic SVS (marked by a positive slope red dotted line) is observed. At yet higher normal stresses no SVW is observed (not shown).

 \item $\mu^{ss}(v)$ for an interface between rough PMMA (a glassy polymer) and smooth silanized glass, data extracted from Fig. 7 of \citet{Bureau2002}. A clear crossover from logarithmic SVW to logarithmic SVS is observed. Note that the smoothness of the substrate (a rigid silanized glass) implies that contact asperities are not continuously formed and destroyed during sliding, and hence that a different mechanism for the crossover (as compared to the one discussed in the text) is involved.

 \item $\mu^{ss}(v)$ for a dry (room-humidity) clay-rich gouge layer, data extracted from Fig. 10 of \citet{Ferri2011}. The curve seems to be velocity-independent at low velocities, followed by a quasi-logarithmic SVS and then a crossover to stronger-than-logarithmic velocity-strengthening. Eventually, a sharp decrease in friction is observed at high velocities.

 \item $\partial \mu^{ss}/\partial\log v$ for calcite (upper panel) and dolomite (lower panel), both data sets extracted from Fig. 2 of \citet{Weeks1993}. A crossover from SVW to SVS is observed ($\partial \mu^{ss}/\partial\log v\!<\!0$ implies SVW, while $\partial \mu^{ss}/\partial\log v\!>\!0$ implies SVS).

 \item The same as panel (d), but for granite. Logarithmic SVW (marked by a horizontal line) and a crossover to stronger-than-logarithmic SVS are observed.

 \item $\mu^{ss}(v)$ for aplite (upper panel) and granite (lower panel). Data courtesy of \textit{Di~Toro, Goldsby, and
  Tullis} (unpublished data, 2013). A crossover from SVW to SVS is observed in both data sets.

 \item $\mu^{ss}(v)$ for Bristol paper, data extracted from Fig. 4 of \citet{Heslot1994}. The curve shows logarithmic SVW at low velocities and a crossover to stronger-than-logarithmic SVS at higher velocities. The inset (linear $v$-axis) shows that $\mu^{ss}\!\propto\!v$ in the SVS regime.

 \item $\mu^{ss}(v)$ for a granular material composed of glass beads under a very low normal stress of $\sigma\!=\!30$ KPa, data extracted from Fig. 2 of \citet{Kuwano2013}. A clear crossover from logarithmic SVW to a stronger-than-logarithmic SVS is observed. The inset (linear $v$-axis) shows that $\mu^{ss}\!\propto\!v$ in the SVS regime.

 \item $\mu^{ss}(v)$ for Sierra White granite in a rotary apparatus, data extracted from Fig. 1 of \citet{Reches2010}. The two curves correspond to two data sets that were selected out of many sets that appeared in the original figure. The lower curve exhibits logarithmic SVW at low velocities, a crossover to stronger-than-logarithmic SVS at higher velocities and eventually SVW at very high slip velocities. The upper curve exhibits a crossover from SVW to stronger-than-logarithmic SVS.\\
\end{enumerate}

Some other works report on SVS and its variability with various control parameters such as the normal stress $\sigma$. For example, \citet{Marone1990} has observed SVS in experiments of simulated fault gouge. The magnitude of SVS varied inversely with the normal stress $\sigma$ and directly with the gouge thickness and surface roughness. In this case, SVS has been associated with granular dilatancy within the gouge layer. In \citet{Kilgore1993}, experiments on bare ground surfaces of Westerly granite have demonstrated a crossover from SVW to SVS at $v_m\!\simeq\!10\ \mu$m/s for a normal stress of $\sigma\!=\!5$ MPa. For higher normal stresses, up to 150 MPa, SVS has not been observed in the range of measured slip velocities (up to $10^3\ \mu$m/s). It is not entirely clear whether SVS did not exist under these conditions or was simply shifted to higher slip velocities.

Velocity-strengthening friction has been also quite extensively discussed in the context of clay-rich fault gouge layers, which often exhibit only velocity-strengthening behavior \citep{Noda2009, Ikari2009}, cf. the data for dry clay-rich gouge shown in Fig. \ref{exp_evidence}c. Additional experimental data for $\partial\mu^{ss}(v)/\partial\log{v}$ in three different types of wet (saturated) clay-rich gouge layers are shown in Fig. \ref{clay}. All of these data sets exhibit stronger-than-logarithmic SVS, roughly consistent with $\partial\mu^{ss}(v)/\partial\log{v}\!\simeq\!c_1\!+\!c_2\log{v}$, where $c_1,c_2\!>\!0$ and $c_2$ is similar in magnitude to $c_1$ or larger. It is not entirely clear whether wet clay-rich gouge can be in fact properly described by the physical framework discussed above. On the one hand, the real contact area $A_r$ may not be an adequate state variable in the context of clay-rich gouge and on the other hand, additional state variables associated with compaction, permeability and hydration may be required. Yet, constitutive frameworks similar to the one discussed above were invoked in the literature to phenomenologically interpret the frictional behavior of clay-rich gouge layers \citep{Ikari2009, Noda2009, DenHartog2012, DenHartog2013}; hence it might be somewhat useful to speculate about this issue here.

First, we note that the relation $\partial\mu^{ss}(v)/\partial\log{v}\!\simeq\!c_1\!+\!c_2\log{v}$, which roughly characterizes the data in Fig. \ref{clay}, seems consistent with Eq. (\ref{eq:RSF}) for $v\!\ll\!D/\phi^*$, where $c_1\!=\!\alpha - \beta$ and $c_2\!=\!-\alpha\beta/f_0\!=\!-\alpha\,b$. This, however, requires $b$ (or equivalently $\beta$) to be negative, as was indeed suggested in \citep{Ikari2009}. This possibility has to be treated as a phenomenological reinterpretation of the aging coefficient $b$, which as such cannot possibly be negative. Furthermore, typical values of $\alpha$ and $\beta$ are much smaller than unity, implying $c_2\!\ll\!c_1$, which does not seem to be the case in the data of Fig. \ref{clay}. Another possible origin of the $v$-dependence of $\partial\mu^{ss}(v)/\partial\log{v}$ can be that $\alpha$ is no longer $v$-independent, in the spirit of the discussion in Sec. \ref{sec:rehology}.

Putting aside the question of the $v$-dependence of $\partial\mu^{ss}(v)/\partial\log{v}$, one can speculate why $\partial\mu^{ss}(v)/\partial\log{v}$ is at all positive in clay-rich gouge layers, i.e. why $c_1\!>\!0$. The first possibility is that the real contact area $A_r$ is not a relevant state variable for these systems in the range of slip velocities probed. This can happen either because $v\!\gg\!D/\phi^*$ in this range of $v$'s, leading to $\log\left(1+D/v\,\phi^*\right)\!\to\!0$, or $b\!=\!0$. Both of these possibilities were discussed to some extent in \citet{Ikari2009}. If the real contact of area is relevant, then Eq. (\ref{logarithmicder}) tells us that $\beta\!=\!f_0\,b$ should be sufficiently small. This of course can be achieved if $b$ is small (the limit $b\!\to\!0$ was mentioned in the previous paragraph) or the background level of the friction coefficient $f_0$ is small. Interestingly, this is indeed the case for the clay-rich gouge systems presented in Fig. \ref{clay}, whose steady-state coefficient of sliding friction was systematically smaller than $0.35$. We finally note that while this discussion of clay-rich gouge systems is speculative, we hope it does shed some light on the variety of behaviors that can emerge from the proposed framework.

All in all, we believe that the diverse experimental data sets presented and discussed in this section imply that the physical picture depicted above should be seriously considered.

\section{Summary and Discussion}
\label{summary}

In this brief note we argue that a steady-state velocity-strengthening behavior might be a generic feature of dry friction over some range of slip velocities. We stress that the emergence of velocity-strengthening is a natural consequence of an experimentally well-established phenomenological picture of dry friction at relatively low slip velocities. In this picture, logarithmic velocity-weakening friction (dominated by the ``rejuvenation'' of contact asperities) crosses over to logarithmic velocity-strengthening friction (dominated by thermally-activated rheology) at a typical slip velocity $\sim\! D/\phi^*$ where the real contact area saturates. We further suggest that logarithmic steady-state velocity-strengthening friction should cross over to a stronger-than-logarithmic velocity-strengthening behavior at a slip velocity $v_0$, typically accompanied by a change in the dominant frictional dissipation mechanism.

The above discussed scenario is expected to hold if $D/\phi^*\!<\!v_0$. However, as $D/\phi^*$ and $v_0$ correspond to different pieces of physics, one cannot exclude the possibility that $v_0\!<\!D/\phi^*$. In this case we expect logarithmic velocity-weakening friction to cross over to stronger-than-logarithmic velocity-strengthening friction at $\sim v_0$. Some examples in Fig. \ref{exp_evidence} seem to support this possibility. Moreover, this behavior is expected to be the generic case in athermal systems (e.g. granular materials), where no thermally-activated rheology is relevant (cf. Fig. \ref{exp_evidence}h).

\begin{figure}[h!]
 \centering
 \includegraphics[width=0.7\columnwidth]{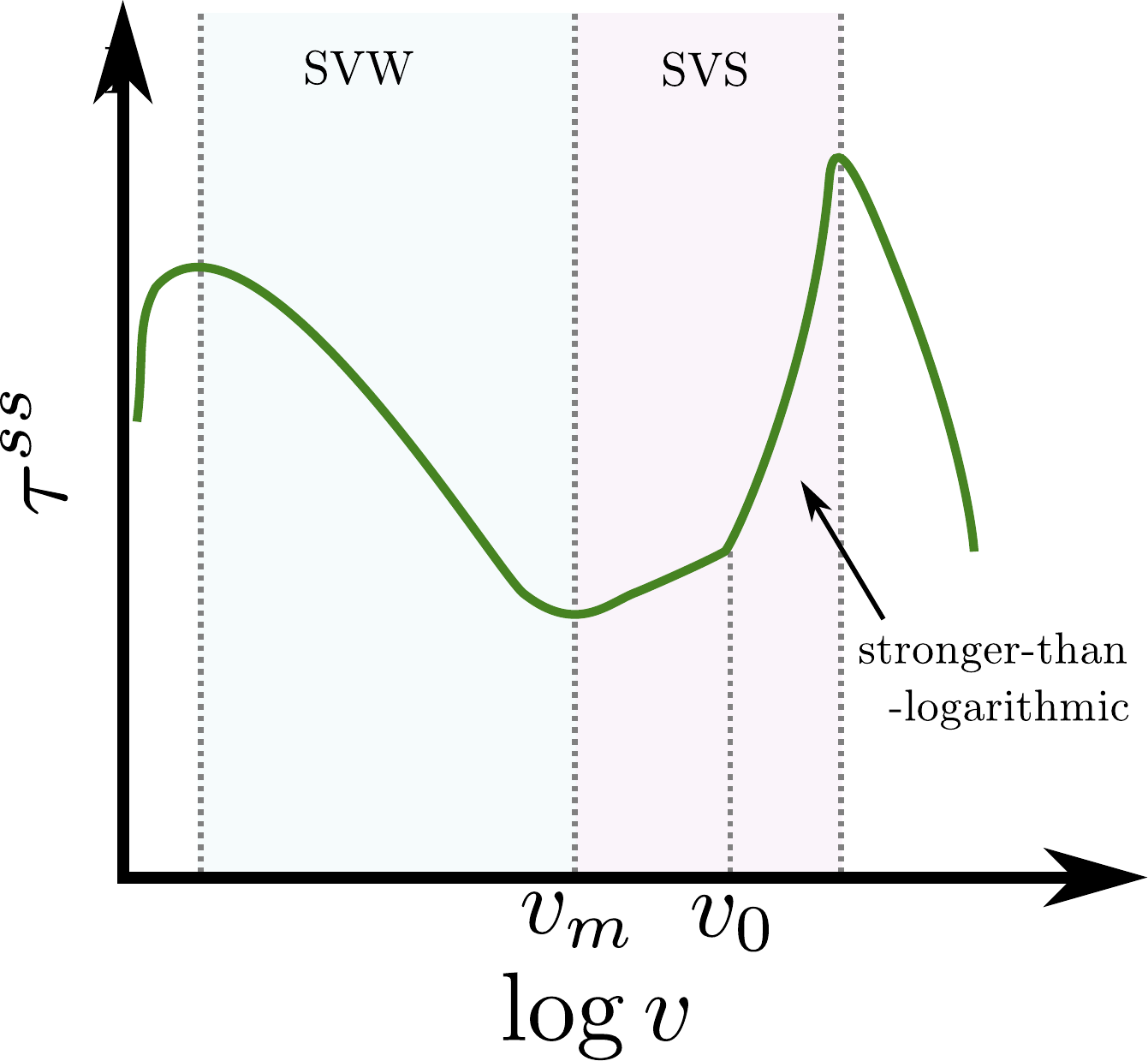}
 \caption{A schematic picture of the steady-state friction law in the case $\alpha\!<\!\beta$ and $v_0\!>\!v_m$. The color code follows Fig. \ref{exp_evidence}. At extremely low sliding velocities friction is velocity-strengthening, governed by plastic creep (not discussed in this paper). At higher velocities, but below $v_m$, friction is logarithmically SVW. At $v\!=\!v_m$ friction becomes logarithmically SVS, and at $v\!=\!v_0$ it crosses over to stronger-than-logarithmic behavior. At even higher velocities, friction might decrease substantially due to thermal weakening effects (not discussed). If $\alpha\!>\!\beta$ the first peak will not exist, and if $v_0\!<\!v_m$ the logarithmic strengthening regime will be absent.}
 \label{MShape}
\end{figure}

We compile a rather large number of experimental data sets available in the literature, directly demonstrating the existence of steady-state velocity-strengthening friction (both logarithmic and stronger-than-logarithmic). These examples cover a rather wide range of materials, including various rocks (e.g. granite and halite), a glassy polymer (PMMA) -- widely used in laboratory experiments -- on smooth silanized glass, a granular material (glass beads), clay-rich gouge layers and Bristol board. We suspect that this behavior is robust and will be observed in many other materials as long as careful steady-state friction experiments cover a sufficiently large range of slip velocities.

We should mention two other aspects of steady-state friction that were not discussed above, but are observed in some of the data sets presented in Fig. \ref{exp_evidence}. First, at extremely small slip velocities one expects friction to be velocity-strengthening due to creep-like plastic flow response \citep{Estrin1996}. This is clearly observed in panel a. At very large slip velocities thermal weakening effects might be operative, leading to significant (sometimes overwhelming) velocity-weakening friction \citep{DiToro2004, Rice2006, Goldsby2011}. This is clearly observed in panels c and i. Combining these features with the previously discussed ones, an M-like friction curve emerges, as schematically shown in Fig. \ref{clay}. Note, however, that in some cases (e.g. the clay-rich gouge layers data in Fig. \ref{clay}) the first peak might be missing. Yet, in all of the wide variety of examples presented, velocity-strengthening friction exists, which is our main point.

The existence of velocity-strengthening friction might have serious implications for various frictional phenomena. While these have not been studied extensively in the literature up to now, we would like to mention here the effect of velocity-strengthening friction on the upper cutoff in seismicity along well-developed faults \citep{Marone1988}, its effect on earthquake afterslip and negative stress drops \citep{Marone1991}, the role played by velocity-strengthening friction in the stability of homogeneous sliding between dissimilar materials \citep{Rice2001}, in facilitating slow slip events \citep{Weeks1993, Kato2003, Shibazaki2003, Bouchbinder2011, BarSinai2012, Hawthorne2013, Bar-Sinai2013pre} and in giving rise to steady-state interfacial rupture fronts under stress-controlled boundary conditions \citep{BarSinai2012}. We hope that the present note will encourage further research in these, and other, directions.

\begin{acknowledgments}
We are grateful to Prof. T. Shimamoto for a constructive and useful set of comments that helped us improve the manuscript. EB acknowledges support of the James S. McDonnell Foundation, the Minerva Foundation with funding from the Federal German Ministry for Education and Research, the Harold Perlman Family Foundation and the William Z. and Eda Bess Novick Young Scientist Fund. EAB acknowledges support of the Erna and Jacob Michael visiting professorship funds at Weizmann Institute of Science.
\end{acknowledgments}

\end{document}